\documentclass{aa}

\usepackage{graphics}

\begin{document}

\title{Classical, non-linear, internal dynamics of large, isolated, vibrationally excited molecules }

\author{R. Papoular
\inst{1}
          }

\offprints{R. Papoular}

\institute{Service d'Astrophysique and Service de Chimie Moleculaire, CEA Saclay, 91191 Gif-s-Yvette, France\\
              e-mail: papoular@wanadoo.fr
             }

\date{Accepted July 31, 2002}

\authorrunning {R. Papoular}

\titlerunning {Intramolecular dynamics of large molecules}

\maketitle

\begin{abstract}
This work reports numerical experiments intended to clarify the internal equilibration process in large molecules, following vibrational excitation. A model of amorphous and oxygenated hydrocarbon macromolecule ($\sim$500 atoms)--simulating interstellar dust-- is built up by means of a chemical simulation code. Its structure is optimized, and its normal modes determined. About 4.5 eV of potential energy is then deposited locally by perturbing one of the C-H peripheral bonds, thus simulating the capture of a free H atom by a dangling C bond. The ensuing relaxation of the system is followed for up to 300 ps, using a molecular mechanics code. When steady state is reached, spectra and time correlation functions of kinetic energy and bond length fluctuations indicate that most normal modes have been activated, but the motion remains quasi-periodic or regular. By contrast, when the molecule is violently excited or embedded in a thermal bath (modelled by Langevin dynamics), the same markers clearly depict chaotic motions. Thus it appears that even such a large system of oscillators is unable to provide the equivalent of a thermal bath to any one of these, barring strong resonances between some of them. This conclusion is of consequence for the interpretation of astronomical infrared spectra.
 
Collateral numerical experiments show that a) relaxation times increase as perturbation energy decreases by spreading through the system; b) energy deposited in the highest-frequency modes does not relax preferentially into the lowest frequency-modes but follows specific paths determined by near resonances and coupling strength; c) energy deposited in the lowest-frequency modes is able to flow up the whole frequency ladder, albeit less easily than in the opposite direction.

\vfill\eject

 %
  \end{abstract}

%
\section{Introduction}
The present research was triggered by the question whether infrared chemiluminescence was possible with large molecules. This question, in turn, arose from the need to interpret the flurry of accurate IR (infrared) spectra delivered by dedicated astrophysics satellites in the past two decades. It is indeed established that the ISM (InterStellar Medium) is interspersed 
with large molecules or small grains (dust), with a wide size distribution (up to 1000 nm), and whose composition is essentially limited to hydrocarbons and silicates. We are interested, here, in that major family of amorphous hydrocarbons which lie so far from any illuminating star that they can only reach a low steady state temperature ($\leq$200 K) when the UV radiative power they absorb equals the IR radiative power they emit. Under such circumstances, the relatively high  IR spectral intensities observed are incompatible with Planck's law at that temperature. A possible alternative excitation mechanism is chemiluminescence, based on the fact that, in such environments, the dominant material, hydrogen, remains in  atomic form, which notably favours interaction with hydrocarbon dust. Such interaction is believed to be the main cause of formation of molecular hydrogen, which is the main constituent of the "clouds" that are detected within the ISM (Duley and Williams$^{1}$).
When a free H atom hits an H atom that is attached to a C atom at the periphery of a dust particle, they may form such a strong bond that the nascent H$_{2}$ molecule can fly away, leaving a dangling bond behind. Upon a later encounter with an H atom, the latter can be captured by the dangling bond to form a long-lived C-H bond. In such an event, an energy of order 4.5 eV (the bond dissociation energy) is deposited \emph{locally} in the molecule. This work addresses the fate of that energy and of the resulting atomic vibrations: what vibrational modes are excited and can they survive energy redistribution long enough for them to finally relax through the relatively slow process of  IR emission?

Of course, IR chemiluminescence is known to occur upon reactions between small molecules (e.g. see Lee and Pimentel$^{2}$). Is this possible with larger molecules as well? and how much larger? More precisely, does statistical equilibrium set in, with equal probability for all states of equal energy (also referred to as complete randomization)? If it does, a temperature can be defined, which scales like the inverse of the number of atoms; then, for a dust particle larger than, say, 100 atoms, the temperature would be so low as to preclude detectable emission of all but the very lowest vibrational frequencies, contrary to observations. If, on the contrary, complete randomization does not obtain, then the motions of the various atoms retain at least some degree of coherence and the concept of temperature becomes irrelevant. The distribution of energy among the various modes is then totally different, being determined by classical ( or quantum, for more accuracy) mechanics: even the highest vibrational frequencies can be excited to a notable extent and IR chemiluminescence becomes detectable.

The RRKM theory of unimolecular reactions (see Gilbert and Smith$^{3}$) indeed assumes thermodynamical equilibrium, and is vindicated by innumerable experimental data. However, these data are mostly about yields of particular unimolar reactions (dissociations) averaged over assemblies of molecules under finite pressure, so that vibrational excitation is effected collisionally and randomly, according to equilibrium thermodynamics. This, according to Freed$^{4}$ , is enough to ensure the validity of all the mathematical predictions of the RRKM theory, even assuming no randomization at all occurs within the molecule! Besides, more recent experiments have shown deviation from statistical behaviour, especially when the energy content is low (see McDonald and Marcus$^{5}$). Other types of experiments, especially those designed in pursuit of "selective chemistry", suggest that, under suitable conditions, the molecule can be prepared in an excited \emph{and coherent} state, which can survive for quite long times (see Zewail$^{6}$). More recently, Williams and Leone$^{7}$ followed the IR-\emph{radiative} relaxation of gaseous naphtalene at very low pressures, after laser excitation of a C-H stretch vibration; they showed that their spectra are not compatible with calculations based on thermodynamic equilibrium. Also, Deak et al.$^{8}$ excited the C-H stretch of nitromethane molecules in a tiny segment of a jet, and observed the prompt ($\leq$2 ps) establishment of  a nearly uniform distribution among other vibrations, which prevailed for a time about equal to the collisional relaxation time of the bath (30 to 50 ps). Afterwards, a new, Boltzmannian, distribution clearly set in, again suggesting that collisions where responsible for this to occur. Similar conclusions were drawn by Stromberg et al.$^{9}$ , working on W(CO)$_{6}$. Time and frequency resolved experiments reveal in fact three cases of IVR: dissipative (energy sharing), restricted, and no IVR at all (see, for instance, Zewail$^{10}$). Correspondingly, molecular motions range from statistical, ergodic, chaotic, to quasi-periodic and periodic, or regular.

Turning to theory, it is known that the normal modes of vibration of a system are orthogonal and, therefore, in principle do not interact. However, they do so in real life, because anharmonicity always provides some coupling between them, albeit weak. This coupling increases in strength as the molecule becomes larger, and the distance between modes correspondingly  decreases in frequency space. There is therefore no doubt that any local disturbance is bound to decay in some way, except, in principle, in the ideal limit of the weakest excitation of the purest normal mode. This type of reasoning led Fermi, in the 30's, to conjecture that a statistical (Boltzmannian) equilibrium  must ultimately set in (see Ford$^{11}$). He therefore initiated numerical simulations of the dynamics of long chains of oscillators, the outcome of which did not, however, confirm his conjecture: hence the Fermi-Ulam-Pasta paradox (Fermi$^{12}$). Long afterwards it became clear that statistical intramolecular energy transfer is not ubiquitous (see Brumer$^{13}$) even with many degrees of freedom; conversely, under suitable circumstances, "chaos" can occur even with a few degrees of freedom. Thus, the size of the system  and the corresponding number of normal modes are not good indicators of fast randomization $^{14}$. 

Finally, a hierarchy of statisticality (stochasticity) and randomness was established, with ergodic, mixing, K and Bernouilli systems in order of increasing randomness$^{15}$. Conditions favourable to chaos and randomness were recognized (see Ford$^{11,16}$): existence and overlapping of non-linear resonances ($\Sigma$n$_{k}$$\omega$$_{k}$$\leq$$\epsilon$),  high energy content, brief random excitations (as in a perfect gas or with billiard balls), etc.
 
In terms of phase space, Boltzmann's ergodic hypothesis implies that all available phase space is uniformly visited by the system's trajectory during the observation time. This leads to the usual \emph{canonical prescription} for deriving time averages of physical quantities. It is found that energy is distributed among the degrees of freedom (here, the vibrational modes) according to Boltzmann's law with a single temperature common to all. While this delivers the least biased estimates of observables \emph{in the absence of enough information on the system}, very many well documented physical systems are not properly described in this way$^{17}$. In these cases, the phase space is full of structures which induce "intramolecular bottlenecks" to the phase flow (see Uzer$^{18}$ for a review). Such systems are found to obey a \emph{restricted} form of ergodicity, \emph{broken ergodicity}$^{17}$, meaning that, in these cases, \emph{statisticality (randomness) obtains only in limited regions of phase space}.

For all these developments, theory can hardly predict which type of internal motion will prevail in systems with more than a few degrees of freedom. Surely, experiment remains the ultimate recourse. However, numerical models of coupled oscillators have been of great help in highlighting the role of different parameters (particularly excitation strength and bond anharmonicity), in assessing the degree of randomness, in interpreting experimental outcomes and in evaluating relaxation times (see Noid et al.$^{19}$ and Thompson and co-workers$^{20}$). With the advent of highly performing chemical software, numerical modeling can now more closely simulate the motion of large, real-life molecules. This is the course followed in the present work to address the problem mentioned above.

The main result of this research is as follows: in the type of structure considered here, and following the recombination of an impinging H atom with a dangling C bond, the energy deposited is ultimately, and on average, roughly equipartitioned among all the vibrational modes. However, chaos does not ensue and the molecule remains in a nearly coherent state, i.e. a coherent superposition of normal vibration modes . A Fourier analysis of the time variations of the total electric dipole moment then provides the spectral distribution of energy, which shows that chemiluminescence is indeed detectable.

The originality of this work does not lie in the method used nor in the conclusion that statistical equilibrium does not always obtain after relaxation. Rather, it is in the combination of very large molecules ($\sim$ 500 atoms), amorphous structure, chemical excitation, no collision at all before complete internal relaxation and very long computation runs (up to 500 ps). Such a combination, suggested by  astronomical conditions, is difficult to achieve in the laboratory and, to my knowledge, was never considered before. Moreover, the outcome of this study may seem counterintuitive to many who are not familiar with the FUP paradox, and therefore may fuel interesting  debate.

\section{The chemical software}

The calculations were made using the Hyperchem package released by Hypercube, Inc. in 1999, and implemented on a desk-top PC equipped with a Pentium (r)II processor (450 MHz) and MMX (TM) technology.
First, the molecule is built on the screen by picking atom after atom in a table of elements. The CC bondings are specified graphically: -, =, $\equiv$. A computational code and options are then selected. The code is then asked to optimize the molecular geometry, i.e. determine the architecture which minimizes the total potential energy, E$_{0}$.
A perturbation is then applied by displacing one or more atoms, or imparting a velocity to them, or creating a new bond (like in  chemisorption). This brings it up to an energy state E*. The subsequent motion is then followed using a Molecular Dynamics routine based on the Born-Oppenheimer approximation: at each calculation step, the nuclei are assumed to be clamped at their initial positions and the code determines the electronic charge distribution, either by solving Schroedinger's equation ((a) below) or empirically ((b) below). Hence the forces acting on the nuclei and the new velocities of the latter, which determine the displacement for the next step. The step length is set at 10$^{-16}$ s, much less than the shortest characteristic period of the system. To save time, the electric dipole moment is computed only once every cycle of 30 dynamics steps, still ensuring correct frequency analysis up to 3000 cm$^{-1}$ .

For a frequency resolution of 30 cm$^{-1}$, a Molecular Dynamics run must span ~1 ps,   i.e. 10000 steps or 300 cycles. However, much longer times are needed for transients to decay. The coordinates and velocity of each nucleus, and hence the potential and kinetic energies are recomputed at every step. The total energy is conserved all along, as it should be, even for the longest calculation. The energy content, or perturbation energy is Ep= E*-E$_{0}$ .
The accuracy and speed of this computation depend on the choice of the code; while most state-of-the-art chemical codes are available with this package, only two of them were used in the present work:

a)	the semi-empirical PM3 code, developed by J. Stewart$^{21}$ along the same line as AM1. The semi-empirical methods use a rigorous quantum-mechanical formalism combined with empirical parameters obtained from comparison with experimental results. PM3 incorporates a much larger number and wider variety of experimental data than AM1. These codes compute approximate solutions of Schroedinger's equation, and use experimental data only when the quantum mechanical calculations are too difficult or too lengthy. This sometimes makes them more accurate than poor ab initio methods, and they are always faster.
PM3 also provides for a direct calculation of the normal mode spectrum, which does not require running the Molecular Dynamics. This is based on the quadratic term of the interatomic potentials and only involves matrix manipulation. The result applies, more generally, to weak excursions from the equilibrium geometry. The spectrum is discrete, i.e. composed of separate vertical bars of length equal to the  "IR intensity", A, of the structure at the corresponding wavenumber. In real life, this must be corrected for broadening effects: thermal fluctuations, rotations, and slight differences between particles of the same family. For large particles, the first two effects are negligible in the astrophysical context.

	The accuracy of the mode wavenumbers has been discussed at length$^{22}$: although it depends on the type of vibration and on the structure, it is considered to be of order 10 cm$^{-1}$ for the C-H stretch and 100-200 for the rest of the spectrum.

	An interesting feature of this code is the possibility of animating the atoms on the screen in any selected normal mode. This is very useful in the process of tailoring the composition and structure of the particle to fit the astronomical spectra as best as possible.

	This code is very much faster than ab initio codes (which are out of the question here). However, for a molecule of 100 atoms, each Molecular Dynamics step requires $\sim$30 s computation time, which amounts to about 100 cycles a day or 1 month for a run of minimum length. This code was used in the present work to determine, by cut-and-try, a structure capable of displaying the main IR features observed in the sky; this structure is shown in fig. 1 (Papoular$^{23}$,$^{24}$). Note that it is 3-D, heterogeneous, disordered, amorphous and closely resembles models of coal$^{25}$ and kerogen$^{26}$, a material also found in meteorites. In this study, we have used larger molecules built up by assembling randomly part or totality of fig. 1 (101 atoms) so as to retain the same chemical characters. Most of the results below refer to the largest of these molecules, designated by LM and consisting of 493 atoms. This is too large to be conveniently handled with semi-empirical codes.

b)	The Molecular Mechanics code MM+.
This is based on the Allinger$^{27}$ MM2 code. Molecular Mechanics codes, like MM2, Amber, Charmm,... approximate the required generalized potential energy surface (force fields) by analytical functions which are deduced from, and subsequently parametrized via, accurate calculations and experimental results. The force field contains terms associated each with bond stretching, angle (in-plane) bending, torsional (out-of-plane) bending, stretch-bend cross terms, twisting, van der Waals  forces and other terms of chemical significance. Anharmonicity is arbitrarily built in various terms; it is cubic for bond stretches.

MM+ uses the latest MM2 parameters and atom types (designed primarily for small organic molecules), complemented, where necessary, by specific calculations. When the "atomic charge" option is selected, the charge on every atom is invariable and determined empirically by the type of atom and bonding. As a result, the vibration frequencies derived from MM+ calculations may be in error by up to 
200 cm$^{-1}$. Moreover, the absolute value of the electric dipole moment is not reliable. However, this code is very much faster than the semi-empirical codes: about 20000 cycles a day for a particle of 100 atoms. It is, therefore, highly useful in clarifying the general trends of the spectra as a function of parameters such as time, energy content, structure, etc. Such trends, which are our main interest here, are not likely to depend upon the exact numerical values of vibration frequencies in large molecules like our's. This code was therefore used extensively for our present purposes; since it is not suited to simulating the hydrogen recombination with the molecule, the latter is approximately simulated by starting with the model particle at rest in its ground state, E$_{0}$, as above . Then, instead of breaking a C-H bond, the bond is simply altered in length and direction on the screen, by cut and try, until the new potential energy of the particle is E*$\sim$E$_{0}$+4.5 eV. The Molecular Dynamics is launched thereafter. Under the perturbation just described, the real molecule is expected to remain in its electronic ground state. Energy is conserved to at least 3 significant digits and much better on average over a few steps. 

The relevance and efficiency of force fields and semi-empirical methods are now well established in spite of their neglect of quantum effects. The rationale behind their extensive use lies in the Ehrenfest theorem (see $^{34}$ for a discussion).

\begin{figure}

\resizebox{\hsize}{!}{\includegraphics{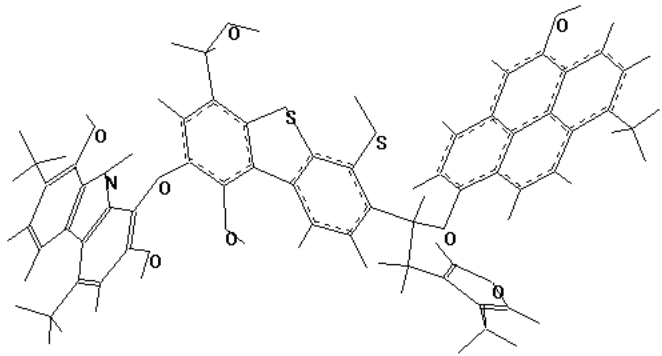}}

\caption[ ]{The optimized structure of the building block (101 atoms) of the reference macromolecule, LM (493 atoms).}
\end{figure}

\section{The approach to steady state}
The quantitative criterion adopted here for stationarity is 

\begin{equation}
$R=M$^{0.5}$$\sigma$/$\langle$E$_{kin}$$\rangle$$
\end{equation}

 where $M=3N-6$ is the number of degrees of freedom for \emph{N} 
atoms, $\langle$E$_{kin}$$\rangle$ is the average total kinetic energy, and $\sigma$, its standard deviation. As a function of time, this tends towards a horizontal asymptote. This is shown in fig. 2 for our largest molecule (LM, 493 atoms). The  relaxation is initially very fast ($\tau$$\sim$0.5 ps) but later looks more like an extended exponential (dashed curve). Ultimately, R$\approx$1 and all the degrees of freedom appear to have been activated.

\begin{figure}
\resizebox{\hsize}{!}{\includegraphics{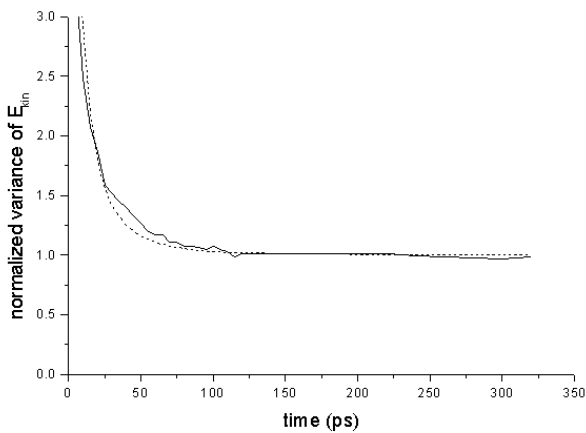}}
%
%
%
\caption{The reduced standard deviation, R=M$^{0.5}$$\sigma$/$\langle$E$_{kin}$$\rangle$  , of the total kinetic
 energy content, E$_{kin}$, of the molecule, as a function of time after local injection of 4.5 eV of chemical energy; $\sigma$ is the std. dev. and M the number of normal modes. The dashed curve is a rough fit with the extended exponential, 1+30/exp(t/0.8)$^{0.4}$.}
\end{figure}

 The code allows us to follow the "diffusion" of energy in time and space, from the perturbed bond into the molecule ; this is done by defining a circle/sphere centered on the captured H atom, "selecting" the atoms within and asking for their total kinetic energy; this procedure is repeated for any desired radius and instant along the calculation. Concurrently, autocorrelation and FFT (Fast Fourier Transform) can be obtained from the time evolution of the position/velocity of any atom and of the length or orientation of any bond, revealing the nature of the dynamics. In this way, it is found that 

1) the deposited chemical (or potential) energy is promptly transformed into stretch ($\sim$3000 cm$^{-1}$)and bend  
($\sim$1500 cm$^{-1}$) vibrations of the central C-H bond; these motions are initially highly non-linear and carry strong harmonics;

2) spatially extended low-frequency vibrations  quickly invade the whole molecule while high-frequency vibrations lag behind because they are more localized in space and preferentially exchange energy between neighbouring atoms;

3) overall, low frequencies grow at the expense of high frequencies, until all modes are activated and rough energy equipartition obtains between all atoms;

4) steady state is essentially established at about 100 ps;

\begin{figure}
\resizebox{\hsize}{!}{\includegraphics{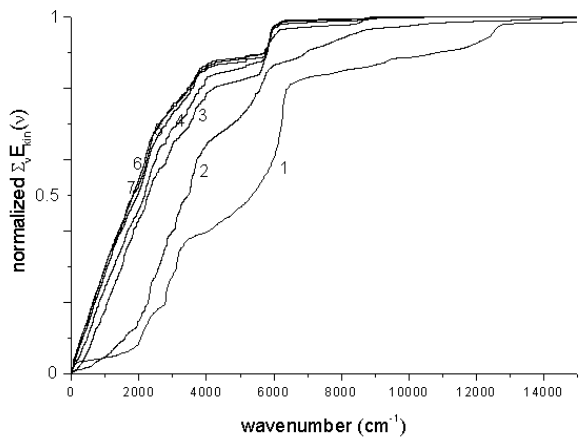}}
\caption[ ]{The relaxation of the cumulative spectral kinetic energy distribution,  $\sum$E$_{kin}$($\nu$) ,  where E$_{kin}$($\nu$) is the Fourier transform of E$_{kin}$(t); normalized to 1 at the highest computed frequency. Curves 1 to 7 were recorded successively after delays of 0.41, 1.6, 5, 55, 71, 121,and 221 ps, respectively, starting at the beginning of the run.}
\end{figure}

One would also wish to know the energy in each normal mode. However, a normal mode analysis is hopeless because of the system size. A less rigorous, but simpler and quite helpful, approach consists in Fourier analyzing the \emph{total} kinetic energy, E$_{kin}$, over an arbitrary period (here  from 0.4 to 16 ps). Because of spectral congestion, this "energy" spectrum is inconvenient for illustrative purposes. It is better to use the \emph {cumulative} spectral kinetic energy distribution, $\sum$E$_{kin}$($\nu$), which is obtained by summing up the spectrum from 0 to $\nu$, the abscissa. The result is drawn in fig.3 for several successive times during the transient; the curves are all normalized to 1 at the highest frequency. Again because of spectral congestion, they seem to be roughly monotonous on this scale, except for the isolated C-H stretch band near 6000cm$^{-1}$. Since E$_{kin}$ is a quadratic form in the coordinates or momenta, the frequencies are, in fact, double the vibration frequencies.  In case of a purely harmonic motion, one also expects sum and difference frequencies. These are washed out, however, in the presence of strong dephasing between all frequency components, which seems to be the case here, making for an easier interpretation. 

Points 3) and 4) made above are synthetically illustrated in fig. 3. It appears that the spectrum is essentially invariable after a few picoseconds, although some "reddening" goes on for a much longer time. It is remarkable, at this point, that the energy initially concentrated in the C-H stretch flows roughly from high to low frequencies: the main relaxation path is not direct to the "bath" frequencies. Neither is it a strict "vibrational cascade" down a ladder of levels: significant amounts of intermediate and low frequency energy are observed at the earliest stage; moreover, the relaxation time roughly increases with decreasing mode frequency. All this was already deduced from experiments on small molecules in liquid environment (see Deak et al. 1999$^{8}$, Sibert and Rey$^{8}$). In the present study, the system is isolated so the energy decay of the C-H stretch modes (5000 to 6000 cm$^{-1}$ in fig. 3) is only partial; as expected from anharmonic oscillators, it is accompanied by line narrowing and blue shift . 

\section{The steady state}
We now take up the detail character of the motion in steady state. There are a number of different features which distinguish the various types of molecular motions (Noid et al.$^{19}$). Most are associated with phase space coordinates; these are well suited to theoretical studies but can hardly be accessed in the present case. More manageable are the methods which involve looking at the frequency spectra or correlation functions of dynamical variables. As noted by Noid et al., the frequency spectrum of a dynamical variable,  for quasi-periodic (regular) motion, consists of a series of sharp lines, while the chaotic frequency spectrum has broad lines forming bands near the location of the sharp lines of the corresponding regular spectrum. The transition from regular to chaotic is also characterized by the continuum between bands becoming less and less smooth and more and more rugged (grassy).
In principle, for a given perturbation, every normal mode shows up, to a smaller or larger extent, in the fluctuations of position and velocity of each atom, as well as of the length of each bond, depending on the type of atom or bond and immediate environment. C-C bonds were found to carry more and stronger spectral lines than C-H bonds. Figure 4 shows the amplitude-FFT of the length of one C-C bond in the LM molecule, more than 220 ps after excitation. The time length of the data sample was 65 ps, so that the frequency resolution is 1 cm$^{-1}$ (with Hanning-filtering). Most lines are indeed about this wide, and roughly retain their (simple) shape but vary in strength to some extent because they exchange energy with distant  modes. Some are so close that they partly blend together. A case in point is the C-H stretch band, which extends from 2920 to 2970 cm$^{-1}$ (somewhat below experimental value, because of imperfections in the model force fields). Since there are about 250 H atoms in this molecule, we expect about that number of C-H str modes, all in that range, but slightly displaced from one another because of differences of environments. Clearly, these modes are close enough to strongly interact and share energy, like \emph{nearly} resonant coupled oscillators. This is indeed what transpires from the comparison of successive FFT's over, say, a 16 ps interval each: the band profile changes but the integrated "intensity" remains essentially constant. This is an example of selectivity in IVR (Felker and Zewail$^{28}$). In this molecule, however, we found no case of strong resonance entailing "thermalization" between the corresponding oscillators, as shown by Ford et al.$^{16}$. By contrast, we did simulate such a resonance by exciting some of the stretches of the highly symmetrical benzene molecule, and observing the subsequent fluctuating energy distribution among the six C-H bonds (the resonant oscillators) in steady state.

\begin{figure}
\resizebox{\hsize}{!}{\includegraphics{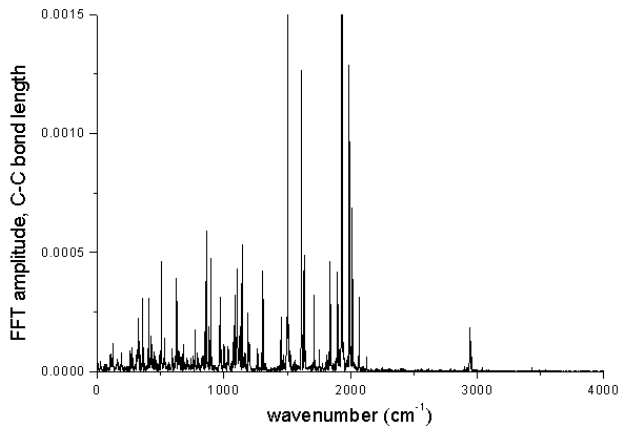}}
\caption[ ]{The frequency spectrum of the fluctuations of the length of one of the C-C bonds, more than 221 ps after excitation. Note the narrowness of the lines, limited by the sample length (65 ps).}
\end{figure}

Overall, spectra like the one in fig. 4 are not grassy but consist of distinct, narrow bands whose positions do not vary notably in time; the Fourier transforms of the corresponding autocorrelation functions deliver identical spectra, frequencywise. The motion can therefore be considered regular overall. This conclusion is confirmed by the autocorrelation functions: e.g., in fig. 5,  for the same C-C bond length as in fig. 4; note the very strong and persistent correlations; the superimposed modulation indicates fluctuations of line intensities due to energy exchanges between oscillators. Similar results are obtained for C-H bonds. Thus, despite heavy spectral congestion, the molecule as a whole does not qualify as a thermalizing bath to any individual mode. Reasons  for this are probably the amorphicity of the structure, which hinders close resonances, and the lack of a "phonon continuum" at the lowest frequencies of the spectrum (see fig. 6 of$^{24}$) as opposed to the case of molecular crystals$^{29}$, for instance.

\begin{figure}
\resizebox{\hsize}{!}{\includegraphics{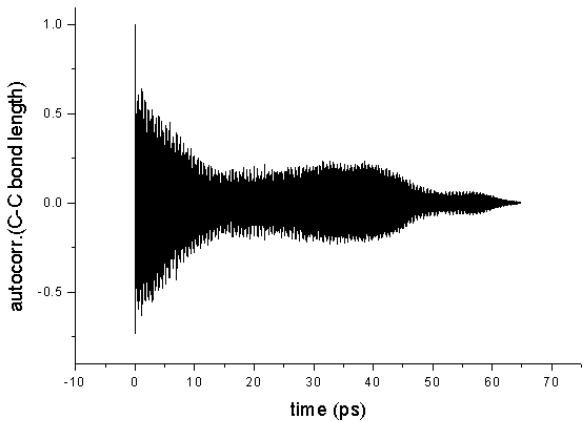}}
\caption[ ]{The autocorrelation function of the fluctuations in length of the same C-C bond as in fig. 4, and during the same time.}
\end{figure}

For our present purposes, it is important to note that this steady state vibrational energy distribution is not very sensitive to the details of the H atom capture or, in terms of the procedure adopted here, to the details of the initial C-H bond perturbation which is meant to simulate that capture, nor to the particular C-H bond that is perturbed, as long as the excess energy thus deposited remains of order 4.5 eV. From a more fundamental point of view, however, it is interesting to observe that, because of the non-linearity of the system, the steady state depends on that excess energy as well as on the type of bond which is perturbed (C-C, O-H, stretch, bend, torsion, etc.). None the less, even if the perturbation consists in altering the torsion angle of a C-O-H group, thus exciting one of the lowest-frequency modes, vibrational energy finds its way up the frequency ladder and most of the normal modes are ultimately activated. Figure 6 illustrates the relaxation  of LM to equilibrium after the torsion angles of six such groups were altered by about 100 degrees. Note the differences between the steady state spectrum reached here (curve 6) and the one that was reached after altering the length of a C-H bond (curve 7, identical with curve 7 of fig. 3): the higher frequencies are distinctly weaker here, and the opposite is true below 2000 cm$^{-1}$. The same trend is observed, but to a lesser degree, if energy is instead initially deposited in a C-C bond. This sensitivity of the final state to the initial perturbation is an indication that statistical (thermodynamic) relaxation did not take place.

\begin{figure}
\resizebox{\hsize}{!}{\includegraphics{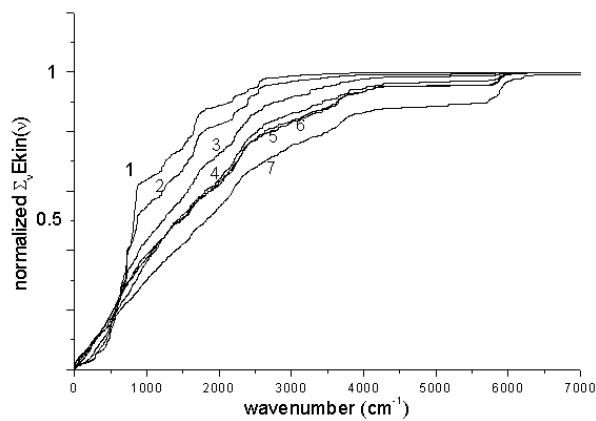}}
\caption[ ]{The relaxation of the cumulative spectral distribution, $\sum$E$_{kin}$($\nu$), after perturbation of the torsion angles of several C-O-H groups, which adds about 0.6 eV to the ground state of the molecule. Spectra 1 to 6 were recorded after 8, 16, 65, 200, 450 and 500 ps, respectively. Spectrum 7
 is identical with 7 of fig. 3 and is added here for comparison. Note the general evolution from left to right by contrast with fig. 3; also, the asymptotic spectrum 6 is poorer than 7 in higher frequencies.}
\end{figure}

It can be shown$^{24}$ that the expected spectral density of IR radiation from the excited molecule is proportional to the product of the square of the frequency with the power spectral density of the dipole  moment. The latter is the FFT of the sum of the autocorrelation functions of the fluctuations of the dipole moment components about their respective means. Finally, these fluctuations can be recorded during a run of the code well after steady state has settled. This procedure was applied to the steady state corresponding to fig. 3 (curve 7) and fig. 4, and the result is displayed in fig. 7. This shows that only a fraction of the vibrational modes is "infrared active", but one can still recognize the C-H stretch bands (aromatic and aliphatic, around 3000 cm$^{-1}$), the in-plane C-H bends ($\sim$ 1000-1200 cm$^{-1}$) and the C-H out-of-plane bends ($\sim$ 750-1000 cm$^{-1}$), and far infrared bands mainly associated with oxygen atoms in various environments$^{23}$. It should be stressed again that the present computations used the MM+ molecular dynamics software, which is known to yield only approximate frequencies and intensities. A more accurate spectrum can be obtained using the PM3 semi-empirical software, albeit on a smaller molecule (100 atoms), but still representative of the present amorphous structure$^{24}$. 

\begin{figure}

\resizebox{\hsize}{!}{\includegraphics{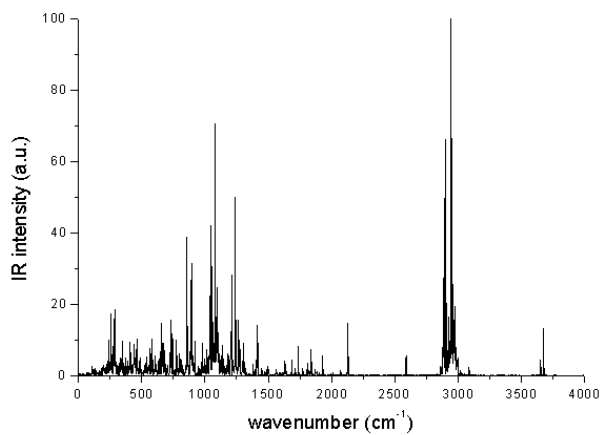}}
\caption[ ]{The infrared spectrum of the isolated molecule LM in steady state after chemical excitation by capture of a H atom.}
\end{figure}

\section{Chaotic motions}
The conclusions drawn above are strengthened by comparison with conspicuously chaotic motions. Such motions are expected when the molecule is immersed in a thermal bath. The Hyperchem package allows the simulation of this situation by means of Langevin dynamics, in which the usual equations of motion are supplemented by terms representing both the friction against the bath molecules and the random changes in position and velocity upon collisions with these molecules (Allen and Tildesley$^{30}$). The user may specify the temperature, T, and the friction coefficient, $\gamma$; the stronger the latter the tighter the coupling of the molecule to the bath. For purposes of comparison with fig. 5, we set $\gamma$=100 ps$^{-1}$ and T=35 K, a temperature at which the energy content of the molecule is about the same as above. Figure 8a shows, under these conditions, the autocorrelation function of the length of the same C-C bond as in fig. 5. The difference is striking: very weak correlation, with no regular pattern.

\begin{figure}
\resizebox{\hsize}{!}{\includegraphics{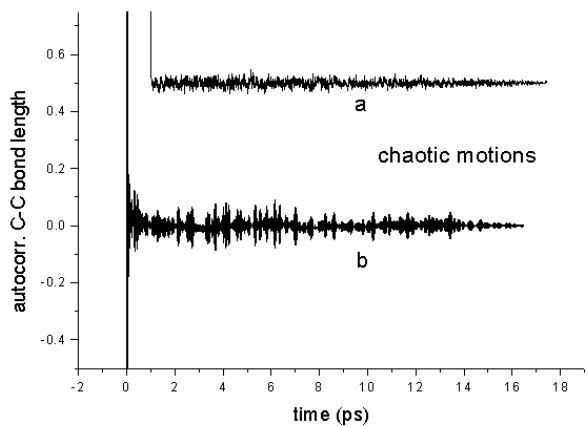}}
\caption[ ]{Autocorrelations of time fluctuations of the C-C bond length for chaotic motions: a) molecule in a thermal bath at 35 K, simulated by Langevin dynamics (curve shifted by (+1,+0.5) for clarity); b) isolated but highly excited molecule (12.5 eV). The autocorrelations are characteristically weaker than in fig. 5. }
\end{figure}

Theoretical studies also show that chaos is bound to set in beyond a total energy threshold, even in an isolated system. To explore this route to chaos, we first ran Langevin dynamics at 3000 K for 0.3 ps. This dumped an energy of 12.5 eV uniformly in the molecule.The latter was then isolated from the bath and molecular dynamics was run in vacuum for 16.5 ps. The corresponding correlation function is shown as curve b in fig. 8, again for the same C-C bond length as above. The weak correlations and highly disordered modulation indicate the partial onset of chaos, but this cannot be carried much further because the molecules dissociate at higher excitation energies. Chaos is much easier to induce that way in a small and/or symmetric structure like benzene; I confirmed this using the same code.

The frequency spectra of these chaotic motions are also distinctly different from the regular spectra: they are very grassy and individual modes are no longer recognizable because they blend together, even at the highest resolution. In the extreme case of Langevin dynamics, all features are drowned in the grass which extends to much higher frequencies. This is apparent from the cumulative spectral kinetic energy distributions in fig. 9, where curves 9a,b correspond, respectively, to 8a,b while 9c is taken from fig. 3, the steady state regular motion. In such graphs, of course, the grass is not apparent because it is integrated out, so that the progression to chaos from c to a is made clearer. 

\begin{figure}
\resizebox{\hsize}{!}{\includegraphics{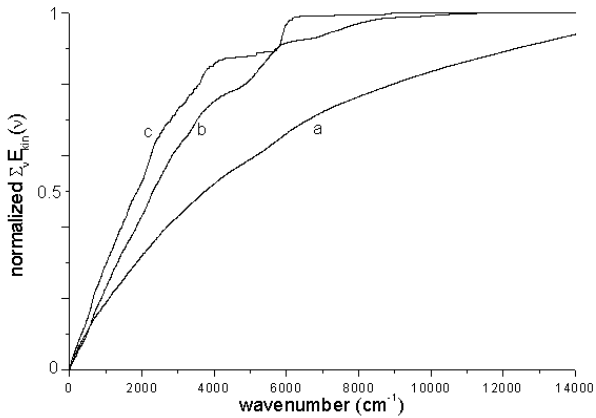}}
\caption[ ]{The cumulative spectral energy distributions a,b and c, corresponding, respectively, to autocorrelation curve a of fig. 8 (Langevin dynamics), curve b of fig. 8 (violently  excited isolated molecule), and curve 7 of fig. 3 (mildly excited isolated molecule). Note the progression from quasi-regular to fully chaotic motion, going from c to a.}
\end{figure}

\section{Conclusion}
Given the size and structure of our system, and the type and strength of the activation process, there is no theoretical way to determine, beforehand, the nature of the subsequent relaxation. The computations performed in this study help drawing a realistic classical model of the dynamics to be expected.

Many markers point to a regular (quasi-periodic) motion in steady state: the steady state depends on the initial perturbation (although not wildly); the spectral lines are narrow; correlations are strong and long lived; the spectral energy distribution (energy spectrum) is very different than for a molecule in a (model) thermal bath; in particular, there is no continuum here. However, this steady state is not \emph{perfectly} coherent: the spectral lines vary \emph{erratically} in intensity (although their short term averages are constant), i.e. there are \emph{random}, reversible, energy exchanges between modes; and the combination frequencies are absent from the energy spectrum, i.e. there are no apparent phase relations between modes.

These results are all understandable in terms of more or less strongly and non-linearly coupled anharmonic oscillators (see $^{11,16}$ and references therein). They fit neatly in the picture of partial chaos or broken ergodicity$^{17}$: the phase point can wander randomly, but only in restricted regions of the energetically available phase space. Such situations occur when the vibrational spectrum is too sparse and/or the energy content is too low for multiple overlaps to occur between non-linear resonances. In the present instance, this is likely due to the limited size, and to the amorphicity and inhomogeneity of the structure.

From comparisons between corresponding classical and quantal motions (see Noid et al.$^{19}$), it appears that quantum chaos does not set in before classical chaos is well established. It seems, therefore, that,  even under the more realistic, but "corresponding", laws of quantum mechanics, the vibrational motions of our activated molecule will not reach statistical (thermodynamic) equilibrium. If that is the case, then the classical spectral intensity (fig. 6) may be interpreted as the ensemble average quantal energy content at the same frequency. One can therefore expect that such molecules, isolated and chemically excited, will ultimately deliver a significant part of their initial energy content in the form of near and mid-IR radiation. 

The question of the upper size limit beyond which these conclusions no longer apply remains open to discussion. It is known that the vibrational energy deposited in a molecular \emph{crystals} will ultimately flow into the \emph{phonon bath}$^{29,31}$. Phonons are very low-frequency ($<$100 cm$^{-1}$) plane propagating waves which can transport energy through the structure. They can interact upon hitting a structural or chemical defect, and exchange energy like perfect gas particles in a vessel, so that thermal equilibrium can set in. But, for the localized vibrations (vibrons), which we have been considering above, to be efficiently converted into phonons, the spectrum of the latter must be dense enough as to form a quasi-continuum to which the Fermi Golden Rule of transitions can be applied. This is only possible if the structure is regular and infinite, like ideal crystals; very irregular structures obviously can hardly support phonons, a fact which translates into very low thermal conductivities and long thermal relaxation times$^{32}$. Indeed, the relaxation times of some glasses at low temperatures have been found to range in the hours$^{33}$. It is likely, therefore, that a grain of our model material will have to be quite large before it can give rise to a phonon spectrum of sufficient density that it constitutes an efficient thermal bath.

1. W. Duley and D. Williams, \emph{Interstellar Chemistry} (Academic Press, London, 1984).

2. Y.-P. Lee and G. Pimentel, J. Chem. Phys. $\bf{69}$, 3063 (1978).

3. R. Gilbert and S. Smith, \emph{Theory of Unimolecular Reactions} (Blackwell Scientific Publ., London, 1990).

4. K. Freed, Faraday Disc. Roy. Chem. Soc. $\bf{67}$, 231 (1979).

5. J. McDonald and R. Marcus, J. Chem. Phys. $\bf{65}$, 2180 (1976).

6. A. Zewail, Acc. Chem. Res. $\bf{13}$, 360 (1980).

7. R. Williams and S. Leone, Astrophys. J. $\bf{443}$, 675 (1995).

8. J. Deak, L. Iwaki and D. Dlott, J. Phys. Chem. A $\bf{103}$, 971 (1999); E. Sibert III and R. Rey, J. Chem. Phys. $\bf{116}$, 237 (2002).

9. C. Stromberg D. Myers and M. Fayer, J. Chem. Phys. $\bf{116}$,3540 (2002).

10. A. Zewail, in \emph{The Nobel Prizes} (Almqvist and Wiksell, Stockholm, 1999).

11. J. Ford, Phys. Rep. $\bf{213}$, 271 (1992).

12. E. Fermi, \emph{Collected Papers}, Vol. II (U. Chicago Press, 1965), p. 978.

13. P. Brumer, Adv. Chem. Phys. $\bf{47}$, 201 (1981).

14. D. Nesbitt and R. Field, J. Chem. Phys. $\bf{100}$, 12735 (1996).

15. J. Lebowitz and O. Penrose, Phys. Today $\bf{26}$, 23 (1973). 

16. J. Ford and J. Waters, J. Math. Phys. $\bf{4}$, 1293 (1963); J. Ford and G. Lunsford, Phys. Rev. $\bf{A 1}$, 59 (1970).

17. R. Palmer,  Adv. Phys. $\bf{31}$, 669 (1982).

18. T. Uzer, Phys. Reports 199, $\bf{73}$ (1990).

19. D. Noid, M. Koszykowski and R. Marcus, Ann. Rev. Phys. Chem. $\bf{32}$, 267 (1981).

20. T. Sewell, D. Thompson and R. Levine, J. Phys. Chem. $\bf{96}$, 8006 (1992); B. Sumpter and D. Thompson, J. Chem. Phys. $\bf{86}$, 2805 (1987) and $\bf{82}$, 4557 (1985).

21. J. Stewart, J. Am. Chem. Soc. $\bf{107}$, 3902 (1985).

22. D. Seeger, D. Korzeniewski, W. Kowalchyk, J. Phys. Chem. $\bf{95}$, 6871 (1991).

23. R. Papoular, Astron. Astrophys. $\bf{362}$, L9 (2000).

24. R. Papoular, Spectrochem. Acta, Part $\bf{A 57}$, 947 (2001).

25. J. Speight, Appl. Spectr. Rev. $\bf{29}$, 117 (1994).

26. F. Behar and M. Vandenbroucke, Rev. Inst. Fr. Petrole $\bf{41}$, 173 (1986).

27. N. Allinger, J. Am. Chem. Soc. $\bf{99}$, 827 (1977).

28. P. Felker and A. Zewail, Phys. Rev. Lett. $\bf{53}$, 501 (1984).

29. D. Dlott, in \emph{Laser Spectroscopy of Solids II}, ed. W. Yen, Topics in Applied Physics vol.65 (Springer-Verlag, 1989), p.167.

30.  M. Allen and D. Tildesley, \emph{Computer Simulation of Liquids} (Clarendon Press, Oxford, 1987), p. 261.

31. V. Kenkre, A. Tomakoff and M. Fayer, J. Chem.Phys. $\bf{101}$, 10618 (1994).

32. G. Srivastava, \emph{The physics of phonons} (Adam \& Hilger,1990).

33. E. Donth, The Glass Transition (Springer, 2001).

34. M. Ben-Nun and R. Levine, J. Chem. Phys. $\bf{105}$, 8136 (1996).
\end{document}